# Deepfakes at Face Value: Image and Authority

JAMES RAVI KIRKPATRICK

*University of Oxford and Magdalen College Oxford*

**Abstract**

Deepfakes are synthetic media that superimpose or generate someone's likeness on to pre-existing sound, images, or videos using deep learning methods. Existing accounts of the wrongs involved in creating and distributing deepfakes focus on the harms they cause or the non-normative interests they violate. However, these approaches do not explain how deepfakes can be wrongful even when they cause no harm or set back any other non-normative interest. To address this issue, this paper identifies a neglected reason why deepfakes are wrong: they can subvert our legitimate interests in having authority over the permissible uses of our image and the governance of our identity. We argue that deepfakes are wrong when they usurp our authority to determine the provenance of our own agency by exploiting our biometric features as a generative resource. In particular, we have a specific right against the algorithmic conscription of our identity. We refine the scope of this interest by distinguishing between permissible forms of appropriation, such as artistic depiction, from wrongful algorithmic simulation.

## 1. Introduction

Deepfakes are synthetic media that superimpose or generate someone's likeness onto pre-existing sound, images, or videos using deep learning (DL) methods.[1] Previously, the production of fake content was accessible only to photo-edited and computer-generated imagery experts. But with recent advances in DL and the Internet, manipulated media is now easier to produce and more widespread than ever before (Cole 2017, 2018). Even non-experts can create deepfakes with open-source programs using publicly available content with troubling consequences. Deepfake pornography often circulates widely on social media before removal, with one recent study claiming that 96% of online deepfake videos is nonconsensual pornography (Ajder et al. 2019). Deepfakes have also been used to misrepresent the speech of politicians and other prominent public figures (Milmo and Sauer 2022; Feiner 2024). This raises important and timely ethical questions about whether the creation and transmission of deepfakes is morally wrong and, if so, why.

---

[1] The term 'deepfake', a portmanteau of 'deep learning' and 'fake', originated in 2017 as the name of a Reddit user who used DL methods to substitute the faces of actors in pornographic films with faces of celebrities. The term is now generically applied to sound, still images, and videos that have been digitally manipulated with the help of DL methods.

Existing accounts of the wrongs involved in creating and distributing deepfakes focus on the psychological or reputational harms to the victims; the epistemic harms to our ability to engage in and trust images and sounds; and societal harms involving the perpetuation of social injustices (Chesney and Citron 2019a, 2019b; Diakopoulos and Johnson 2021; Fallis 2021; Fletcher 2018; Franks and Waldman 2019; Harris 2022; Laas 2023; Matthews 2022; Meskys et al. 2020; Öhman 2020; Silbey and Hartzog 2019; Spivak 2019; Rini 2020; Rini and Cohen 2022; de Ruiter 2021; Story and Jenkins 2023). However, while harm plays an important role in explaining the wrong of many deepfakes, these accounts are incomplete insofar as they do not explain how deepfakes can be wrongful even when they neither cause harm nor set back any other non-normative interests.[2]

To address this issue, this paper identifies a neglected reason why the mere creation of deepfakes can constitute a serious moral wrong. We defend the following thesis:

> **The Authority Interest View.** Deepfakes are morally wrong in part when they violate the subject's interest in having authority over who does what with their image, likeness, or other identifying aspects of their identity. Specifically, we have an interest in having authority over the governance of our biometric identity and a specific right against its algorithmic conscription. According to this view, deepfakes are morally wrong when they violate our authority to determine the provenance of our own agency, treating our identity as a generative resource for synthetic media.

Our proposal is motivated by familiar discussions of *bare* or *harmless wronging*, situations involving "an action which is a wronging but not in virtue of its being an action against any human [non-normative] interest" (Owens 2012: 8).[3] Consider, for example, how a trespasser seems to wrong you merely by trespassing on your property, even if he does not harm you or damage your possessions (Ripstein 2006: 218). This example of a bare wronging poses a problem for any attempt to ground the wrong of trespass solely in terms of the harms it can cause, even if trespassing normally harms the property owner. We argue that analogous cases potentially pose a problem for existing harm-based accounts of the wrong of deepfakes. In turn, we argue that such cases motivate the Authority Interest View. Ultimately, we argue that the relevant authority interests are partly grounded in our fundamental interest in authorship: our interest in being the sole origin of the actions and performances attributed to our identity.

---

[2] Here and throughout, we follow Owens (2012) in distinguishing between *normative* and *non-normative interests*. A non-normative interest is an interest that one has independently of whether one has a right to them or whether others have obligations to respect them. For example, you have a non-normative interest in being liked by others because this is good *for you*, even though others do not necessarily *wrong* you if they dislike you. In contrast, a normative interest are interests that do or should ground rights, duties, permissions, and obligations. For example, you have an irreducibly normative interest in bodily integrity, that is, in being able to control whether a specific act counts as a wrong against you by giving or withholding consent. This interest matters for its own sake and helps justify the duties of others not to harm or violate you.

[3] For further discussion of bare wrongings, see Feinberg (1990) and Ripstein (2006).

This paper is structured as follows. Section 2 outlines the dominant view that deepfakes are wrong only if they cause or constitute harm to someone in some kind of way. Section 3 introduces the problem of 'bare wronging' and argues that harm-based accounts are incomplete because they cannot explain the wrongness of deepfakes created for private use. Section 4 argues that other existing accounts are also unable to provide a complete account of the constitutive wrong involved in such cases. Section 5 presents our proposal and argues that it explains the bare wrong involved in creating deepfakes. We also clarify the scope of our view by distinguishing between permissible forms of image appropriation, such as artistic depiction and private mental imaginings, from wrongful algorithmic simulation. Section 6 concludes.

## 2. Deepfakes and their harms

A common view is that it is wrong to create and distribute deepfakes because of the harms that this can cause.

> **The Harm-Based Account.** Deepfakes are morally wrong when they cause or constitute harm to someone in some kind of way.

The Harm-Based Account is appealing in its simplicity, although this characterisation leaves open several matters of substance. First, it does not take a stand on whether morally wrong deepfakes must constitute harm or merely cause harm. Second, it does not take a stand on what kinds of harm are morally relevant. Third, it does not specify the ways in which deepfakes must harm to count as a moral wrong. Nevertheless, the Harm-Based Account is a powerful tool that captures intuitions about the wrongness of deepfakes.

To see this, let us consider two broad classes of harms to which deepfakes may give rise. First, deepfakes may be wrong because of the *moral* harms they cause (Chesney and Citron 2019a,b; Diakopoulos and Johnson 2021; Fletcher 2018; Franks and Waldman 2019; Meskys et al. 2020; Öhman 2020; Silbey and Hartzog 2019; Spivak 2019; de Ruiter 2021; Rini and Cohen 2022; Story and Jenkins 2023). Chesney and Citron (2019a) identify several harmful uses of deepfake technology to individuals or organisations, including exploitation, blackmail and intimidation, and false portrayals of speech or endorsement. In turn, these uses can lead to financial extortion, reputational damage, and serious psychological and emotional harm. Deepfakes also pose societal harms, including distorting public discourse, manipulating elections, and undermining trust in democratic institutions (Chesney and Citron 2019a; Diakopoulos and Johnson 2021).

A second class of harms is *epistemic* harms: harms that negatively affect the epistemic status of an agent or group of agents. Deepfakes have the potential to cause epistemic harm, not only by increasing the number of false beliefs we have through misrepresentation, but by reducing the extent to which we can rely on or trust in audiovisual media for knowledge (Rini 2020; Fallis 2021; Harris 2022; Matthews 2022; Laas 2023).[4] Recordings serve as 'epistemic backstops' by correcting false testimony and disincentivising dishonesty, yet deepfakes undermine this regulatory function (Rini 2020). In turn, this could have epistemically deleterious effects for democracy and our dependence on recordings for testimony. Deepfakes not only increase the number of false beliefs through misrepresentation and deceit, but they also reduce the number of true beliefs by sowing the seeds of doubt about the legitimacy of recordings (Fallis 2021). Epistemic and moral harms should be distinguished from one another. For while being made epistemically worse off may make one morally worse off, say, by negatively affecting one's welfare, one is not necessarily made morally worse worth simply in virtue of being made epistemically worse off.

The Harm-Based Account adequately explains why the creation and distribution of a wide range of deepfakes are morally wrong. For example, the wrongness of non-consensual pornographic deepfakes can be partly explained in terms of potential psychological and emotional harm, as well as reputational damage and further repercussions for one's career and employment. Furthermore, the wrongness of deepfakes of political figures making controversial statements or false endorsements can be explained in terms of the societal harm such deepfakes causes to our political and democratic institutions, as well as the more general epistemic harms they may cause.

The Harm-Based Account also explains why certain deepfakes are not morally wrong. For example, deepfakes can be used educate about how such technology can be used to manipulate images and videos. Such deepfakes may not necessarily harm the subjects of the content if it is made clear that the images have been manipulated. Even if these videos are clipped and reproduced outside the original context, potentially causing misinformation once divorced from their educational message, it is the *distribution of the edited material* that is morally problematic, not the creation of the deepfakes in the first place. The pedagogical benefits of bringing the concept of deepfakes and their potential to cause epistemic harms to public attention could outweigh the epistemic risk that the deepfakes are taken as veridical.

---

[4] For arguments that these claims are overstated, see Harris (2021), Atencia-Linares and Artiga (2022), and Habgood-Coote (2023).

## 3. The problem of bare wronging

While many deepfakes are harmful for exactly the reasons outlined above, these harms do not exhaust the reasons why deepfakes can be morally wrong. The Harm-Based Account assumes that the set of wrongful acts is entirely contained within the set of harmful acts. However, there is a class of cases, so-called *harmless* or *bare wrongings*, where an act is wrong despite causing no harm. To see this in the context of deepfakes, consider the following case originally introduced by Öhman (2020):[5]

> **Pornographic Deepfake**. Suppose that A uploads a number of self-depicting images and videos onto the Internet, thus putting them into the public domain. B then uses these pictures as inputs to a Deep Learning algorithm to create a pornographic deepfake of A. Let us further suppose that (i) the technology used by B guarantees that A can never find out about the pornographic deepfake, and (ii) it is impossible to distribute the deepfakes to anyone else, or indeed, for anyone else to come to know about its existence. (Öhman 2020: 134)

Öhman writes: "I claim, the moral intuition of most people is that B is doing something wrong" (2020: 134). We take it that most people would agree with Öhman that it is wrong for B to create the deepfake of A in this manner, even if it is impossible to distribute it or for anyone else, including the victim, to come to know of its existence. Nevertheless, some readers may worry that the motivating intuition about the creation of pornographic deepfakes for exclusively private use is not sufficiently strong to vindicate the idea that creating such deepfakes necessarily violates a morally valuable interest. For one thing, in extant discussion of other cases of harmless wrongs, it is intuitively obvious that the harmless version of the offence is morally objectionable. The most vivid example is the controversial case of "harmless rapes" (i.e., a rape of an insensate victim who never discovers the offence, and never experiences physical or psychological harm or trauma because of the offence).[6] Such cases clearly involve serious moral wrongdoing, and so explanations of the wrongness of rape can rely on the incontestability of this intuition to motivate the idea that rape violates rights or interests that do not depend on harm. In contrast, the intuitions about the wrongness of creating of deepfakes for exclusively private consumption are arguably less resolute. Consequently, it is worth motivating the intuition that the creation of private pornographic deepfakes is morally wrong.[7]

---

[5] Öhman originally introduces this case to explore the moral contrast between deepfake pornography and other actions we do not necessarily find morally objectionable, namely, sexual fantasies. In contrast, we are only interested in explaining the wrongness of deepfakes.
[6] For discussion of so-called "harmless" rape, see Gardner and Shute (2000) and Greasley (2023).

The creation of a pornographic deepfake is an intervention in the representation of another person. Unlike a mere sexual fantasy, which remains an internal, unexpressed mental state generated by the observer's own imagination, a deepfake involves the unauthorized extraction and repurposing of the subject's identity-bearing data. It is a simulacrum of the subject that misrepresents them as engaging in sexually explicit acts they have not actually committed, in a mode of representation to which they have not consented. In this way, the mere creation of a deepfake constitutes a usurpation of agency and normative standing. It appropriates the subject's identity as a raw material, conscripting their likeness by imposing a fictionalized sexual script they did not authorize. This is not a morally neutral act of sexual imagination, but an expressive act of subordination: rather than treating the subject as an agent to be respected, their identity is used as a generative resource to be manipulated in service of sexual gratification without due regard for their autonomy, authority, or control over how they are represented. As Langton (1993) argues in her analysis of pornography as subordination, such expressive acts can wrong others without producing perceptible harm: by bypassing the subject's consent to determine their own representation, the creator effectively enacts a form of subordination that denies the subject their legitimate authority of refusal.

Importantly, this remains true even if the deepfake is never shared. The wrong lies not only in the consequences of distribution, but in the provenance of the creation itself. The creator treats the subject's identity as a resource they are entitled to commandeer. While distribution would cause harms such as reputational damage and psychological stress, these are not the only relevant moral considerations. Even when the deepfake remains private, it expresses an attitude of domination and disregard. The subject is used as a tool for another's gratification; a prop that is denied their autonomy and treated as lacking self-determination; a means to an end determined solely by a creator who acts as if they are entitled to subordinate another's will in accordance with their own fantasies. It says: your likeness is mine to reconfigure; your interests do not matter. Thus, *Pornographic Deepfake* is not morally neutral simply because it is hidden from view. Its wrongness lies in what it expresses and enacts: an erasure of autonomy, a denial of consent, and a symbolic overreach into another's identity. It challenges the moral limits of what we may do with the images and identities of others, even in private.

We find these considerations to be compelling reasons for thinking private deepfakes are morally objectionable, though admittedly, some may still not share this intuition, particularly when compared to other bare wrongings like harmless rape where consensus is broad. Unlike those cases, private deepfakes may appear to some as a technological extension of a sexual fantasy rather than a genuine moral violation. At this point, we must acknowledge a genuine intuition stalemate, and so the aim of this paper is more circumscribed: it offers an account that explains the normative stakes for those who already share the intuition that such acts are wrongful, identifying the specific interests at play when we view deepfakes as a usurpation of authority.

---

[7] Thanks to an anonymous reviewer for pushing us to motivate the central intuition here.

Granting that private deepfakes are wrong, can the Harm-Based Account explain why this deepfake is wrong? Not obviously. The stipulated conditions in *Pornographic Deepfake* forestall any arguments that the deepfake is wrong because of factors based on A's personal well-being or risk to her reputation or future employment. It is far from obvious that A can be harmed by something that causes her no physical or mental damage, that she cannot come to know anything about, and that does not affect her in any tangible way. There is no risk that the deepfake could be used to steal A's identity, to blackmail or intimidate her, or to affect her reputation, career, or employment. One might try to argue that A is harmed because B comes to see her in certain lights, perhaps as a sexual object (Harris 2021: 13387 ft. 18). Indeed, one might argue that B could end up sexually objectifying A even if B never ends up viewing the deepfake, and so never ends up seeing A as a sexual object. Or suppose that A is just one of many potential victims, randomly selected by a lottery algorithm that takes caches of publicly available photographs to generate deepfakes so that B never has any knowledge of whom the deepfake is about. One might still argue that the very act of producing the deepfake of A might be sufficient to endorse her sexualisation. However, while we should certainly admit that the sexual objectification of someone oftentimes amounts to a *wrong* against them, it is difficult to classify B's sexually objectifying A in *Pornographic Deepfake* as B *harming* A in the welfarist sense, especially since A never encounters B and B never interacts with anyone with regards to A. There is no sense in which B's sexually objectifying A results in physical or psychological damage to A, nor does it affect A's welfare in any other tangible way. The pure harm theorist must conclude that there is no wrong. Yet our intuition that B has done something impermissible persists. This suggests that the wrongness is *constitutive*, not causal. B has violated a normative boundary – A's authority over their own representation – rather than setting back A's welfare interests.

For these reasons, the Harm-Based Account cannot provide a full account of the wrongs involved in creating deepfakes.

## 4. Gender oppression, control, and other considerations

How can we explain the moral wrongness in the mere creation of a pornographic deepfake like *Pornographic Deepfake*, if it does not cause any identifiable harms? A natural place to start is with a suggestion made by Öhman himself. He argues that the intuition that B is doing something wrong by creating a deepfake of A rests partly on the role that pornography plays in a broader social system of gender oppression.[8] The idea is that deepfake pornography is enabled by a complex network of male consumers, producers, and misogyny that systematically reduces women to sexual objects. Given that pornographic deepfakes are typically highly gendered, Öhman urges us to accept that the phenomenon is inseparable from the systematic degradation of women. This explains why it might seem more impermissible to create and distribute a deepfake of a famous actress than of a male politician, even if both are produced for the purpose of sexual pleasure. For just like regular pornography, deepfake pornography has the potential to reinforce gender oppression by eroticising relations of gender inequality (cf. Dines et al. 1998; Eaton 2007):

> It is true that both [the famous actress and the male politician] have interests in not having the film made. But when understood through the macro lens of gender inequality—e.g. the technology, the producer, and [the victims] as parts of a larger system, as opposed to merely two arbitrary individuals—these interests differ in legitimacy. Thus the ethical significance of what seems private, and local, lies in the political and social system in which it takes place. (Öhman 2020: 137)

We do not deny the force of this explanation. Öhman is likely correct that the vast majority of actual deepfake pornography constitutes a specific form of gendered violence that functions to uphold patriarchal oppression. However, while this account explains the *severity* and *systemic function* of the wrong in many cases, it does not isolate the *constitutive wrong* of this use of the technology itself. It explains why the deepfake is bad for women, but not why the deepfake is a wrong in itself. To see why, suppose that the victim in *Pornographic Deepfake* is a cisgendered, heterosexual, white, Western, rich male. We suspect that most people would agree that this individual would be wronged by a pornographic deepfake under conditions (i) and (ii) above, just as the actress would be. If our intuition holds that this individual is still wronged, then the wrong-making feature cannot be *solely* derived from gender oppression.

How else might we try to explain the moral wrongness in *Pornographic Deepfake*? Let us consider another strategy, proposed by Story and Jenkins (2023), who argue that non-consensually distributing intentionally non-veridical representations about someone is inherently a *pro tanto* wrong on the grounds that it violates their right that their social identity not be tampered with. This right is grounded in our interest in being able to exercise autonomy over our social relations with others. One way of grounding an autonomy-based right concerning one's social identity stems from de Ruiter's work on deepfakes, who argues that disseminating non-consensual deepfakes should be thought of as a violation of a person's right to digital self-representation, that is, as a kind of 'digital persona plagiarism' (de Ruiter 2021). The idea is that a person's self-image and likeness is a central part of their self-conception, and digital representations play a central role in shaping one's social identity. The creation and distribution of non-consensual deepfakes violates a person's right to be the primary determinant of their social identity, to "present them in ways that disregard their will and go against their sense of self" (de Ruiter 2021: 1327). Another way that Story and Jenkins consider grounding their account, drawing on the work of Rachels (1975) and Marmor (2015), is in terms of privacy rights, specifically in the idea that we have interests in being able to control what sorts of relations we have with others. The idea here is that, in order to exercise autonomy and control over one's social relations, it is important to be able to control the information others have access to about oneself. According to this account, disseminating non-consensual deepfake pornography is wrong because it undermines the subject's ability to exercise control over their relationships with others by undermining their ability to control what information about themselves others have access to.

---

[8] See, also, Rini and Cohen (2022), who argue that deepfake pornography should be seen as *virtual domination*, a form of sexual objectification aimed against specific women.

However, while these strategies get something importantly right in explaining the wrong of deepfakes in terms of the centrality of one's self-image and self-conception to shaping one's social identity, these strategies only explain what is wrong with the *dissemination* of non-consensual deepfake pornography, and not the mere *creation* of it. More specifically, we cannot say that someone has had their social identity tampered with in the mere creation of a deepfake, if that deepfake is never disseminated. Nor can we say A's interests in having control over their relationship with others has been violated, since no new information about A has been disseminated and so A's interests in controlling the flow of information have not been violated. Consequently, while these strategies go some way in explaining what is wrong with the dissemination of deepfakes, it fails to explain what is wrong with the mere creation of them.[9]

A further limitation of Story and Jenkins's and de Ruiter's strategies is that they cannot explain the following case of non-consensual dissemination of deepfake pornography.

> **Pornstar's Deepfake**. A pornographic film actor has fallen victim to a deepfake that depicts them in a qualitatively indistinguishable way from one of their actual scenes. The deepfaker has employed a body-double to reenact one of the actor's scenes, producing a frame-by-frame duplicate, and then used DL technologies to replace the body-double's face with the actor's.

---

[9] Indeed, Story and Jenkins ultimately conclude that "it is not obvious that the person who makes their own deepfake pornography and guards it solely for their own consumption does anything inherently wrong" (Story and Jenkins 2023: 56). We disagree with Story and Jenkins on this point, since we think that A is wronged by the mere creation of the deepfake. But we agree that A's self-image and self-conception play an important role in explaining the wrongs involved here, although the proposal we develop in Section 5 will be importantly different.

This deepfake is veridical in terms of its informational content: it depicts an event that has actually occurred. Given that the original pornographic film is already widely available by the actor's consent, does this mean the actor is not wronged by the creation and dissemination of the deepfake?

We think not. It seems to us that the actor is still wronged, despite the accuracy of the representation. However, existing accounts struggle to explain why. Story and Jenkins focus on the subject's interest in controlling *social identity* and *informational access*. Yet, in this case, the social identity presented is accurate – the actor really is a pornographic film actor who performed in a scene just like in the deepfake – and the information is already public. If the wrongness depended solely on controlling *what* is seen *by whom*, there would be no wrong here. The wrongness must therefore lie elsewhere.

To understand the nature of this wrong, we must distinguish between the *content* of a representation and its *provenance* or *authorship*. Consider the following analogy:

> **Forged Cheque**. Suppose A owes B £50. A intends to pay, but before A can write the cheque, B steals A's chequebook, forges A's signature to write a cheque to himself for exactly £50, and cashes it.

Even though the content of the check accurately reflects the debt that A owes B, and B gets the money that he is owed, B has undeniably wronged A by forging the cheque. Furthermore, this wrong is entirely separable from whether the forged cheque is used to settle A's debt, since B still seems to have wronged A even if he falls victim to a fatal accident that prevents him from cashing the forged cheque. We think that what explains why B wrongs A is the fact that B has usurping A's authority to be the source of the transaction. A has an interest in having the normative power to authorize the transfer and B bypassed that power.

The same reasoning applies to *Pornstar's Deepfake*. Even though the content of the deepfake is veridical in the sense that it depicts a scene that the actor has already performed, the deepfaker has 'forged' the performance. By using the actor's biometric data to generate the video, the deepfaker usurps the actor's standing to authorize the use of their body in that way. The wrong lies not in revealing information already in the public domain, but by bypassing the actor's authority to be the author of their own digital presence. Because Story and Jenkins focus on our interests in controlling information rather than our interests in having normative authority, their account does not capture this distinction.

Finally, let us consider whether the wrongness of generating deepfakes can be explained in terms to privacy violations (Harris 2019). The right to privacy, as it is typically understood following Prosser (1960), is divided into four categories: unreasonable intrusion upon another's seclusion and private affairs, public discourse about embarrassing private facts about another person, publicity which places another in a false light, and appropriation of another's name or likeness for one's own advantage. Neither the first nor second division clearly apply in *Pornographic Deepfake* and *Pornstar's Deepfake*.

For the mere creation of the deepfake does not consist in an unreasonable intrusion upon another's seclusion or private affairs, since the images and videos on which the DL algorithm is trained were put in the public domain by the subject. Similarly, in the present case, the deepfake does not consist in nor would lead to public discourse about embarrassing private facts about another, since the deepfake cannot be disseminated and hence cannot become public knowledge. More generally, non-veridical deepfakes by definition do not represent facts and so cannot constitute an embarrassing private fact about someone (unless its mere creation constitutes such a fact). Third, while it is possible to think of deepfakes generally as involving publicity which places another in a false light, the setup of the case guarantees that there is no such publicity, and so we cannot appeal to the third division to explain the wrongness of the case. Interestingly, there is a second challenge here to articulate exactly how one is placed in a false light by *veridical* deepfakes that are accurate representations of events that have actually happened, but which are misleading insofar as they purport to be archival documentation of the actual event. Finally, while deepfakes certainly amount to the appropriation of another's name or likeness, it is far from clear whether deepfakes amount to appropriation *for one's own advantage*. While some deepfakes are made for the creator's own advantage, such as CEO fraud or false endorsements, many deepfakes are made to intimidate, harass, or simply amuse at another's expense. It is not clear that this constitutes a relevant advantage for the purposes of privacy. For these reasons, it is doubtful whether the wrong of deepfakes, and similarly the right to one's self-image, can be properly captured in terms of one's right to privacy.

## 5. The Authority Interest View and The Right Against Algorithmic Conscription

In this section, we argue that the creation of the deepfakes in *Pornographic Deepfake* and *Pornstar's Deepfake* are moral wrongs against their subjects because they undermine the subjects' interests in having authority over who does what with their image. In particular, we defend the following hypothesis:

> **The Authority Interest View.** Deepfakes are morally wrong in part when they violate the subject's interest in having authority over who does what with their image, likeness, or other identifying aspects of their identity. Specifically, we have an interest in having authority over the governance of our biometric identity and a specific right against its algorithmic conscription. According to this view, deepfakes are morally wrong when they violate our authority to determine the provenance of our own agency, treating our identity as a generative resource for synthetic media.

*5.1. Bare wrongings and normative interests*

The central reason why extant accounts fail to explain the wrongness in *Pornographic Deepfake* and *Pornstar's Deepfake* is that they are instances of *bare wrongings*. A wrong is a *bare wronging* when it consists in the breach of an obligation that does not set back any non-normative interest. For example, someone who secretly breaks a promise to avoid photographing a person wrongs them, even if no harm occurs and neither the promisee nor anyone else will ever find out (Foot 2001: 47–51). Similarly, in the case of harmless trespass, the trespasser wrongs you even though there is no sense in which they harm you by setting back your non-normative interests or by causing you injury (Ripstein 2006). How can we explain the wrongs involved in breaking a promise or committing trespass that does not result in any harm?

Here is one promising account of bare wrongs. Owens (2012, 2019) explains cases of bare wrongings in terms of our *normative interests*: interests which we have in controlling certain features of the normative landscape, such as the obligations and commitments we have and make to one another, our own and other's rights and responsibilities, or the information that can permissibly be made available to each other. We have general interests in having the authority to oblige, permit, or forbid people to do certain things distinct from any interest in being benefited or harmed by people doing or not doing those things. Furthermore, it is good for us to have authority over these features of the normative landscape for its own sake. Following Owens (2012, 2019), let us call these interests in the rights and obligations we control *authority interests*.

We can explain the bare wrong of harmlessly breaking a promise if promising serves our interest in having the right and power to create and abolish promissory obligations by declaration, and doing so serves our interest in having the power to ensure that certain acts constitute wrongs. The promisee obliges the promiser not to take his photograph by extracting a promise from her not do so. This creates a normative boundary. By promising not to take the promisee's photograph, the promiser places herself under an obligation not to do so, thereby making that action wrong even if it does not harm him until he releases her from the promise. The fact that promising has a binding effect that creates obligations to act in certain ways can be explained if it matters to us for its own sake that we have the power over these aspects of the normative landscape, specifically, the ability to create such obligations.[10] Similarly, the bare wrong of harmless trespassing can be explained if it matters for its own sake that we have authority over who can make use of our property (Owens 2019). In general, we have interests in who is permitted and who is forbidden to make use of our private property because having this authority matters to us for its own sake. This is a normative interest that is independent of any non-normative interest we may have, such as in reducing the probability of trespass. Consequently, this account explains the normative significance of promise-breaking and harmless trespass even when no non-normative interests are at stake.

---

[10] Of course, most breaches of promises are not simply *bare* wrongings, since such breaches often violate the promisee's non-normative interests, like being reliant on the promisor acting in certain ways. But according to the authority interests approach, these harms are secondary wrongs that are downstream from the primary wrong arising from the violation of the promise as such.

*5.2 Defining the authority interest and the right against algorithmic conscription*

We argue that individuals have an interest in having authority over who does what with our image.[11] More specifically, we have an interest in having authority in controlling who is permitted to do what with our biometric identity and controlling the generative resource that our physical features constitute. In turn, these interests ground a right against algorithmic conscription.

An authority interest over one's image refers to the power that individuals have to make decisions regarding the use of their own image, likeness, and other aspects that are central to their identities. When it comes to one's image – such as one's physical appearance as captured in photographs, videos, or other media – this authority includes the right to decide if and when one's image can be used, displayed, or shared, the right to determine the contexts in which one's image is presented or used, and the right to control whether one's image is used in ways that accurately represent their true actions or intentions.

In the age of AI, our face and voice are no longer just static audio/visual properties. They are data-rich inputs that can be used to generate new, synthetic performances. We have a specific authority interest in determining whether our biometric data is used as raw material for such generation. This authority includes the right to decide if and when one's biometric identity is conscripted to perform actions one has not willed. This is not a blanket right to prevent any depiction of us, but a right to govern the use of our biometric identity as a generative source. As we will see below, this distinguishes the wrong of deepfakes from ordinary depictions (like drawings), which interpret the subject, but do not conscript their actual identity to manufacture presence.

There is one area where the idea that individuals have overriding authority interests over how their image or likeness is used is not only widely accepted, but also instituted in law, namely, Right of Publicity and related privacy laws. For example, under U.S. Right of Publicity laws, individuals have varying degrees of control over the commercial use of their names, images, likenesses, and other identifying features of their identity. This protects against the misappropriation of an individual's name, likeness, or other aspect of their identity from being used in advertising or merchandise for commercial benefit. This principle was affirmed, for example, by the Supreme Court in the seminal case *Zacchini v. Scripps-Howard Broadcasting*, which involved a broadcaster wanting to film and televise a performer's 'human cannonball' routine without his permission. The Court recognised the performer's right to publicity and rejected the broadcaster's First and Fourteenth Amendment defences, noting that their decision was to not only ensure compensation for the performer, but also to provide "an economic incentive for him to make the investment required to produce a performance of interest to the public" 433 U.S. 562, 576 (1977). The importance of right of publicity laws lies not only in ensuring that a celebrity gets properly compensated for the use of their likeness; it is also about the right to control how a celebrity's image or identity is commercialised, or as demonstrated in *Waits v. Frito-Lay*, Inc. 978 F.2d 1093 (9th Cir. 1992), whether they are used in the commercialisation of a product or service at all.

---

[11] One way of making sense of this idea is to think of our image or likeness as a kind of personal property. For example, Sturino (2023) argues that non-consensual deepfakes are morally wrong because they violate individuals' property rights, namely, the right that individuals have over their own likeness as a form of private property. Since we think there is good sense to be made of the value of private property in terms of authority interests (cf. Owens 2019), our claim that authority interests also explain the bare wrongings involved in the mere creation of certain deepfakes complements Sturino's argument.

But while the issues surrounding our rights over our image and likeness relate to the right of publicity, they are not fully reducible to it. The right to publicity is typically grounded in the right to prevent the unauthorized commercial use of one's identity. It makes actionable the use of a person's name or likeness without their consent, though this doctrine is usually applied when a person's name or likeness – usually a celebrity's – is used in a commercial context to promote or advertise a product or service. Given this historical focus on broadly commercial contexts, the right of publicity is usually grounded in economic claims. Consequently, the limited scope of right of publicity laws cannot explain our broader interests over *non*-commercial uses and abuses of our identity. Such interests generate duties and obligations on others to not use our image or likeness in ways for which we have not granted permission, regardless of whether we stand to lose out economically.

Our goal here is to provide a more general account of our authority interests over our image. According to the present account, the authority interest over one's biometric identity stems from the deeper importance of authorship. As we noted earlier, our image and likeness are direct extensions of our identity. In the public sphere, we constitute ourselves through our appearance; our face and voice are the primary mechanisms by which we interact with others and make our agency manifest in the world. We often work hard to shape how we are perceived, curating certain aspects of ourselves in personal, social, and professional spaces to reflect who we are. This process of self-presentation is not merely a matter of vanity or reputation management; it is an exercise in autonomy.

We have a fundamental interest in ensuring that the figure presented to the world as "us" actually corresponds to our own will. This is what we mean by an interest in authorship: the interest in being the sole origin of the actions and performances attributed to our identity. When others manipulate our image by simulating us, they do not merely show us in a false light: they tamper with a mechanism of self-presentation. This explains why this authority interest is distinct from an active desire to control or curate our image. Even a person who has little interest in managing their public image has a fundamental interest in being the sole author of their actions, in determining who is permitted to do what in their name. A deepfake severs the link between the agent and the act. This inflicts a specific kind of normative injury: it forces the subject to "speak" or "act" without their will. When a deepfake appropriates a person's face to represent them as performing an act, it violates this interest in authorship.

Therefore, the authority interest protects not just our non-normative interests (e.g., reputation), but the integrity of the process by which we constitute ourselves in the public sphere. Just as we have an interest in the integrity of our physical body, we have an analogous interest in the integrity of our digital body, the data-rich map of the self that allows us to appear to others. To lose authority over this is to lose a constitutive part of one's agency in the digital age.

*5.3 Authority interests and deepfakes*

Having outlined the Authority Interest View, let us see how it explains the wrongs involved in the creation of the deepfakes in *Pornographic Deepfake* and *Pornstar's Deepfake.*

In *Pornographic Deepfake*, the deepfake is created without the knowledge or permission of the individual whose image is being manipulated and appropriated. This is a direct violation of their right to authorise and control the governance of their biometric identity. It does not matter that the images used by the algorithm are in the public domain or that they have voluntarily placed there by the subject, since making an image public does not grant blanket permission for any use. By generating synthetic media of a subject, the creator undermines the subject's authority to determine the provenance of their own agency. The deepfake conscripts the individual's identity to perform acts that they did not will, constituting a wrong regardless of whether it damages their reputation, relationships, or public perception.

In *Pornstar's Deepfake*, the wrong is more subtle. While the actor has previously consented to appearing in a pornographic film, they have not consented to their biometric identity being used to generate new performances, even if those performances are qualitatively identical to the original. The creation of the deepfake ignores the specific boundaries of the actor's consent. Here, the mere creation of deepfakes denies the subject any agency or control over the production of their image, which comes apart from any benefit to others or potential detriment to the individual. The victim loses their authority over their representation: they are no longer the arbiter of their own image or identity.

One might object that because the content in *Pornstar's Deepfake* is veridical, there is no "false impression" or deception, and thus no wrong. However, this objection confuses content with provenance. Even if the deepfake does not deceive the view about facts, it misattributes agency. Recall the analogy in *Forged Cheque*. If B forges A's signature to write a cheque for a debt A actually owes, B still wrongs A. B has usurped A's standing to be the author of the transaction. Similarly, the deepfaker usurps the actor's standing to authorize the use of their image. The wrong is not in the created content, since this is already public, but in the unauthorized conscription of the subject's identity to generate the media. The deepfaker treats the actor's faces as a tool they are entitled to wield, thus bypassing the actor's authority to be the sole author of their own actions. This also explains why distributing deepfakes with a caveat or disclaimer about the mode of creation does not solve the moral problem. A disclaimer might mitigate against wrongful depiction, but admitting to creating a deepfake does not remedy the wrongful image appropriation, just as admitting to a forgery does not make the forgery permissible.

*5.4 The scope of the authority interest*

A central challenge for the Authority Interest View is the worry that it proves too much. In particular, one might worry that by positing an overriding authority over the use of one's image, the view does not recognize permissible uses of our image that do not require consent.[12] Consider the following cases:

> **Haircut**. A person takes a photograph of a stranger to a hair stylist to indicate a desired hairstyle.
>
> **Oil Painting**. A hobbyist uses a public photograph of a celebrity as a reference image for a realistic oil painting.
>
> **Satire**. An animator draws a crude caricature of a politician for a show that represents them in embarrassing and non-veridical ways.
>
> **Yearbook**. An individual looks at a publicly available yearbook photograph and uses it to stimulate private sexual imagination.

In these cases, an image of the subject is used without their permission, sometimes in ways to which they may object, but that nevertheless do not intuitively seem to wrong them. If the Authority Interest View holds that *any* unauthorized use of one's image constitutes a bare wrong, then the view is implausibly broad. To defend the view, we must provide principled distinctions that explain why the creation of deepfakes is wrong, while these ordinary uses remain permissible.

We have been motivating the idea that we have authority interests over our biometric identity as a generative resource. This is not a blanket right to prevent resemblance or reference. Rather, it is a specific right against the algorithmic conscription of our identity. We shall argue that this allows us to distinguish between permissible depictions and wrongful simulations.

When a customer references a photograph for a haircut, a hobbyist paints a portrait, or an animator draws a cartoon, they are engaged in depiction. The resulting representation is an interpretation of the subject, mediated entirely by the agent's own skills and perception. The subject is the referent, but the agent is the executor. Crucially, the subject's authority interest is not violated here because the normative source of the representation is the creator's own labour. The authority interest protects the subject's right to determine the provenance of their own agency, to ensure that they are the sole source of any performance presented as their own. In an oil painting or caricature, the work is an expressive act *about* the subject; the representation does not purport to be an action by or performance of the subject. Since the artist does not conscript the subject's biometric identity to act as the generative source of the image but rather reinterprets and mediates it through their own agency, the subject's authority remains intact. In general, we do not have the authority to block others from interpreting or referencing our public appearance, only to block them from simulating our presence.

---

[12] Thanks to two anonymous reviewers for pushing us to clarify the scope of the Authority Interest View in this way and for suggesting the following cases.

In contrast, deepfakes involve simulation. Unlike mere depiction, which involves the interpretation of the subject through a mediator, simulation processes the subject. DL algorithms utilize the subject's biometric data as raw material to generate synthetic media. This constitutes a violation of the authority interest because it falsifies the provenance of agency. The wrong here is not a matter of confusion; it is not that an observer might be tricked into thinking the performance is real. Rather the wrong is normative: the creator has usurped the subject's standing as the sole author of their digital presence. This explains why technical measures like watermarks or the fact that the file remains private do not resolve the moral problem. A watermark may prevent an audience from being deceived, but it does not undo the unauthorised conscription of the subject's identity as a generative resource. Just as forging a signature remains a wronging even if the bank recognizes it as a fake, the algorithmic simulation of an agent is a usurpation of their authority regardless of whether the artifact is 'discovered' or whether a disclaimer is attached. A deepfake eliminates a visible distance between the artist's canvas and the subject's body; it does not interpret the subject, but conscripts their identity to execute a synthetic performance.

We must also distinguish between private mental acts and public fabrication. One might worry that claiming authority over one's image would prohibit private sexual imagination. However, authority interests govern the interpersonal normative landscape: acts that can be regulated by shared rules. Private mental states or events do not constitute a "use" of an image in this interpersonal sense. This distinction holds even if one uses a public photograph to aid one's imagination. In these cases, the viewer uses the image merely as a stimulus for a private thought. They do not manipulate the image to create a simulacrum. In contrast, even the creation of a private deepfake involves external manipulation. It processes the subject's biometric data to create a stable artifact that exists independently of the creator's mind. This artifact constitutes a *simulacrum*: an externalized representation that claims the subject's agency. The creation of this artifact falls within the interpersonal landscape because it involves the manipulation of the subject's digital presence in the external world. The wrong arises not from who sees it, but from the fact that the subject's identity has been conscripted into a concrete unauthorized form.

Thus, the Authority Interest View avoids the charge of overreach. By anchoring the wrong in algorithmic conscription rather than mere resemblance, we preserve the moral space for artistic interpretation and private thought, while protecting the subject from having their digital identity hijacked. This account vindicates the intuition that creating private deepfakes can constitute a moral wrong, one that cannot be captured by harms nor dismissed as a mere variant of drawing or imagination.

*5.5 Secondary harms and beneficial uses*

Having established the primary wrong in creating deepfakes, we can explain its relationship to harm. We have argued that violation of the authority interest constitutes a bare wrong. While this normative interest does not reduce to our non-normative interests, it is not wholly independent of them either. According to the present account, the harms that deepfakes cause are *secondary wrongs* against their subjects. Just as the promiser may cause psychological harm to the promisee if he were to find out his photograph has been taken against his request, or the trespasser could have caused property damage through their actions, the creation and distribution of deepfakes has the potential to cause a variety of significant harms. But just as the harms caused by breaking a promise or trespassing amount to secondary wrongs that stem from the primary wrong of violating someone's authority interests, the variety of harms that deepfakes can cause are also secondary. This is not to say that those harms are any less serious. Rather, it highlights that they are downstream consequences of the violation of the subject's authority.

Our account also explains why the creation of deepfakes is not inherently morally wrong; rather it is wrong when unauthorized. Just as you can grant someone permission to use your personal property, even when it would harm you, you can also consent to a deepfake, even if its creation and distribution would harm you. For insofar as you can consent to the harmful effects of actions and exculpate the agent of that action from any wrong against you, you should be able to consent to the harmful caused by a deepfake and so exculpate its creator from any wrongdoing against you. This is explained on an account that locates the wrong of the creation and distribution of deepfakes in the violation of your interests to have authority over what others do with your image. By granting certain permissions, you release others from the obligations they have to act or not to act in certain ways against you. This allows us to recognise the beneficial uses of deepfake technologies. For example, deepfakes can be used to artificially regenerate the voice of people who are left unable to speak due to degenerative illnesses like ALS (Chesney and Citron 2019a; Meskys et al. 2020). They also have important pedagogical roles, such as making manifest their potential to mislead and providing new and engaging learning resources for students (Erduran 2024). Finally, another example of the benefits of deepfakes, which has widely been embraced by Hollywood, is to allow film stars to appear younger than they actually are, or even to 'revive' dead actors onscreen (Mogensen 2021). So long as the relevant permissions have been granted, the authority interest is respected and the deepfakes can be used in healthcare, educational, and commercial settings. These points are important to bear in mind when it comes to legislating against deepfake technology.

Finally, the Authority Interest View extends to other image violations that do not involve the use of AI, such as non-consensual intimate images (NCIIs), such as 'revenge pornography, 'up-skirting', and voyeuristic filming. The wrong of NCIIs is often framed in terms of privacy violations or harm. However, we argue that the constitutive wrong is the denial of the individual's basic right to decide the scope of their exposure, to decide how their most intimate images are used. NCIIs expose parts of our bodies or our private moments in ways we have not agreed to, even if we have exposed these parts of our body or intimate moments in other settings. What distinguishes a consensual sharing from an NCII is not the content of the image, but the validity of the authorization. The victim may have granted permission in one context (e.g., to a partner) but not the other (e.g., to the Internet). By creating or sharing intimate imagery beyond this boundary, the perpetrator usurps the subject's authority to determine the provenance and audience of their own intimacy. Having the authority to share intimate imagery with others is something which is of value, and it matters to us to have an interest in controlling who may permissibly view that imagery. This also explains why sharing *consensually-made* intimate imagery to people who are not the intended viewers amounts to a serious moral wrong, as well as why viewing such imagery is wrong: the viewer is consuming the subject's image in a context to which the subject has not explicitly consented, thereby participating in the violation of the subject's boundary.

## 6. Conclusion

Image violations are not new. What is new is the extent to which, thanks to modern technology, one can easily engage in the violation and misappropriation of another's image or likeness. In this paper, we provide a principled basis for the idea that such violations are not merely distasteful or harmful, but they also constitute a wrong. This wrong is grounded in our interests in having authority over who does what with our image. This account fruitfully explains the wrong of a wide number of deepfakes, as well as other image violations that do not make essential use of AI. The issue lies not with the harm that these image violations cause, although these harms are oftentimes serious and constitute additional grounds for complaint. Instead, the issue lies in our basic interest in having authority over our image.

Understanding the wrongness of deepfakes and other forms of non-consensual intimate imagery in terms of a violation of authority over one's image highlights why legislation must prioritise protecting authority interests. Laws should recognise that the core issue is not just the harm caused, but the fundamental disrespect for personal autonomy involved in these violations. Proper legal protection would affirm that everyone has the right to decide who does what with their image, especially when it involves intimate or private aspects of their identity.[13]

---

13 We should like to thank Gulzaar Barn, Paul Billingham, Joseph Bowen, Caroline Emmer De Albuquerque Green, James Laing, Ingrid Locatelli, and Luise Müller, and two anonymous reviewers for this journal. We should also like to thank the audiences at the Oxford–Berlin Colloquium on AI Ethics held at the Institute for Ethics in AI, University of Oxford and the Digital Humanities & Artificial Intelligence conference at the University of Reading for helpful comments and discussion on the ideas in this paper.

**Statements & Declarations**

**Funding**: The author declares that no funds, grants, or other support were received during the preparation of this manuscript.

**Competing Interests**: The author has no relevant financial or non-financial interests to disclose.